# Directional Asymmetry in Motion Aftereffects and the Role of Eye Vergence in 3D Motion Perception

**Authors:** D. Zhang

**Affiliation:** Communication University of China, Beijing，China

## Abstract

Motion aftereffect (MAE) offers valuable insights into the mechanisms underlying motion-in-depth (MID) perception. This study investigates two critical aspects of MAE in depth: (1) the potential directional asymmetry between motion toward versus away from the observer, and (2) the effect of induced eye vergence on MAE magnitude. We conducted two experiments using random dot stereograms (RDS) to isolate the interocular velocity difference (IOVD) mechanism. In Experiment 1, we compared MAE magnitude following adaptation to motion-toward versus motion-away stimuli with a static fixation point. In Experiment 2, we introduced a fixation point oscillating in depth to induce vergence eye movements during adaptation and testing. Our results revealed a directional asymmetry in MAE strength, with motion-toward adaptation producing stronger aftereffects than motion-away adaptation in Experiment 1. When eye vergence was induced in Experiment 2, this pattern was reversed, with motion-away adaptation yielding stronger MAEs. These findings suggest an important interaction between adaptation direction and eye vergence state in MID perception, highlighting the complex integration of retinal and extra-retinal signals in the visual system's processing of motion through depth.

## 1. Introduction

The visual system's ability to process motion in three-dimensional space is a remarkable feat that relies on multiple binocular cues. Motion aftereffect (MAE), a phenomenon where prolonged exposure to motion in one direction causes a subsequent illusory motion perception in the opposite direction, has proven to be a valuable tool for investigating the neural mechanisms underlying motion perception (Mather et al., 1998; Huk et al., 2001; Harris et al., 2007).

Two primary binocular mechanisms contribute to motion-in-depth (MID) perception: changing disparity over time (CD) and interocular velocity difference (IOVD). The CD mechanism tracks the temporal changes in binocular disparity of objects moving in depth (Cumming & Parker, 1994; Gray & Regan, 1996; Nefs et al., 2010), while the IOVD mechanism computes differences in velocity between the left and right retinal images (Brooks, 2002; Rokers et al., 2008; Czuba et al., 2010).

Recent studies have provided evidence that the IOVD mechanism can generate stronger MAEs in depth compared to the CD mechanism (Czuba et al., 2011; Sakano & Allison, 2014). However, it remains unclear whether there is a directional asymmetry in the strength of these aftereffects based on adaptation direction (toward vs. away from the observer). Previous research has noted anisotropies in stereoscopic vision, such as enhanced sensitivity for crossed versus uncrossed disparities (Mustillo et al., 1988), suggesting the possibility of similar asymmetries in dynamic motion processing.

Another crucial factor in MID perception is the contribution of eye vergence. While retinal signals provide essential information about motion direction, extraretinal signals from eye movements also play a significant role in how we perceive motion through depth (Harris, 2006; Nefs & Harris, 2007; Welchman et al., 2009). However, few studies have systematically examined how induced vergence eye movements might influence adaptation to motion in depth and the resulting MAEs.

In the present study, we address these gaps by investigating two key questions:

1. Is there a directional asymmetry in the magnitude of MAE in depth based on adaptation direction (toward vs. away from the observer)?
2. How does induced eye vergence during adaptation affect the strength and characteristics of MAE in depth?

To isolate the IOVD mechanism, we employed random-dot stereogram (RDS) stimuli with specific manipulations to minimize contributions from the CD mechanism. In Experiment 1, we compared the magnitude of MAE following adaptation to either motion-toward or motion-away stimuli with a static fixation point. In Experiment 2, we introduced a fixation point that oscillated in depth, inducing vergence eye movements while maintaining the same adaptation conditions.

By examining these questions, we aim to provide insights into the complex interactions between motion adaptation, directional asymmetries, and oculomotor signals in the perception of motion through three-dimensional space.

## 2. General Methods

### 2.1 Observers

Three experienced psychophysical observers (males, aged 20-29 years) participated in both experiments. All observers gave informed consent and were naïve to the specific hypotheses of the study. The study was conducted in accordance with the Declaration of Helsinki. Inclusion criteria were: monocular visual acuity better than 10/10 (measured using a decimal scale chart), good binocular stereo vision, no history of functional or organic ocular pathology, no use of medication that might interfere with oculomotor performance, and no visual complaints prior to the experiment.

All observers underwent extensive training to ensure reliable perception of motion in depth before participating in the main experiments. Each observer completed approximately 2,184 trials across both experiments (approximately 4 hours total testing time).

### 2.2 Apparatus

Stimuli were presented on a 3100 lm DLP-link 120Hz 3D (U310W) video-projector from NEC with a native spatial resolution of 1280×800 (WXGA) converted to 16:9 format. The viewing distance was 70 cm, chosen to ensure that vergence cues (i.e., absolute disparity) provided appropriate distance information for MID perception (Howard, 2008; Neveu et al., 2013).

The DLP technology offered minimal ghosting effects due to its projection system. The device was synchronized with active 3D glasses (Eyes3Shut Purple Two) featuring customized shutters with very dark blocking states (>1/1000) to prevent ghosting. Observers used a joystick to report perceived motion direction. All stimuli were generated using Visual Studio 2010 (C#) and OpenGL.

### 2.3 Stimuli

Random-dot stereograms (RDS) were displayed against a gray background (164 cd/m²). Each RDS contained 80 dots: 40 white dots (392 cd/m²) and 40 black dots (6.9 cd/m²).

Each dot subtended 0.15° in diameter with anti-aliasing to achieve precise pixel positioning. The dots moved within a volume defined by a disparity range of ±0.6°.

The RDS was presented in an annular configuration with inner and outer radii of 3° and 7°, respectively. A white fixation cross (subtending 0.4° with a line thickness of 0.08°) was presented in the center of the display throughout the experiment.

The RDS contained both signal dots and noise dots. Signal dots moved coherently in the same direction (either toward or away from the observer) at a constant monocular velocity of 0.6°/s. To prevent tracking of individual dots and to ensure continuous motion, each signal dot was assigned a random initial position within the ±0.6° disparity volume and a random lifetime between 0-250 ms (Czuba et al., 2011).

To avoid potential biases in dot density distribution, we implemented a wrapping procedure where dots that reached the disparity limits (±0.6°) were repositioned to the opposite depth plane while maintaining their motion direction and speed. This ensured that the overall dot density remained uniform throughout the stimulus volume, eliminating any potential static disparity cues that could indicate motion direction.

Noise dots were assigned random motion directions (toward or away) with speeds and lifetimes equal to or less than those of signal dots. The initial positions of noise dots were randomly distributed throughout the stimulus volume, similar to signal dots. To create a smooth transition at the edges of the disparity volume, we implemented a linear contrast ramp that reduced dot visibility as they approached the disparity limits, effectively "blending" them with the background.

## 2.4 General Procedure

Both experiments employed a motion nulling paradigm (Blake & Hiris, 1993; Czuba et al., 2011) to measure the strength of MAE in depth. In each trial, observers reported the perceived direction of motion-in-depth (toward or away) for test stimuli with varying levels of motion coherence. The percentage of "toward" responses was plotted as a function of motion coherence to generate psychometric functions, from which we derived the point of subjective equality (PSE) as a measure of MAE strength.

Each experiment consisted of three sessions: unadapted (baseline), adaptation to motion-toward, and adaptation to motion-away. The unadapted session always occurred first, followed by the two adaptation sessions in counterbalanced order across observers. A minimum rest period of two days was required between sessions to minimize carryover effects.

For the unadapted session, each trial consisted of the following sequence: (1) a 1.25 s inter-stimulus interval (ISI) with only the fixation cross visible; (2) a 1 s presentation of the test stimulus (binocularly correlated RDS with varying motion coherence); and (3) another 1.25 s ISI during which observers reported the perceived motion direction. For adaptation sessions, each trial followed this sequence: (1) a 4 s adaptation period with anti-correlated RDS moving consistently in one direction (either toward or away); (2) a 1.25 s ISI; (3) a 1 s test stimulus; and (4) a 1.25 s ISI for response. An initial 100 s adaptation period preceded the first trial to build up the adaptation effect.

Test stimuli varied in motion coherence, defined as the percentage of signal dots relative to the total number of dots. Positive coherence values indicated motion-toward, while negative values indicated motion-away. In the unadapted session, six coherence levels (±5%, ±30%, ±50%) were tested. In the adaptation sessions, we expanded this range to ten levels (±5%, ±30%, ±50%, ±80%, ±95%) to capture potentially larger adaptation effects.

## 2.5 Data Analysis

For each observer and condition, we fit a cumulative logistic function to the proportion of "toward" responses as a function of motion coherence:

$$F_L(x; \alpha, \beta) = \frac{1}{1+exp^{-(\frac{x-\alpha}{\beta})}} \qquad (1)$$

Where $x \in (-\infty, +\infty), \alpha \in (-\infty, +\infty), \beta \in (-\infty, +\infty)$, α corresponds to the threshold: $F_L(x = \alpha; \alpha, \beta) = 0.5$, it determines the overall position of the curve along the abscissa. Individual psychometric functions were fit for each observer using Psignifit Toolbox 3.0 (Fründ & Haenel, 2011) for Python 2.7, with bootstrapping (2000 repetitions) to estimate confidence intervals for the fitted parameters. We then averaged the derived parameters across observers to obtain group results.

The strength of MAE was quantified as the shift in the PSE (the $\alpha$ parameter) from the unadapted baseline condition. A negative shift indicated that more motion-away signal dots were needed to null the MAE (consistent with a motion-toward aftereffect), while a positive shift indicated that more motion-toward signal dots were needed (consistent with a motion-away aftereffect).

# 3. Experiment 1: Directional Asymmetry in Motion Aftereffect

## 3.1 Methods

Experiment 1 was designed to investigate whether the magnitude of MAE in depth differs depending on the direction of adaptation (toward vs. away from the observer). The fixation cross remained stationary at zero disparity throughout the experiment.

Each observer completed 72 trials (12 trials × 6 coherence levels) in the unadapted session (approximately 4.2 minutes) and 120 trials (12 trials × 10 coherence levels) in each adaptation session (approximately 15 minutes each).

## 3.2 Results

### 3.2.1 Baseline Sensitivity to Motion Direction

In the unadapted condition, we observed a directional asymmetry in observers' sensitivity to motion-in-depth. Figure 1 shows the proportion of correct responses for motion direction discrimination as a function of coherence level, separately for motion-toward and motion-away. At low coherence levels (±5%), observers showed significantly better performance for motion-toward than motion-away ($p = 0.0026$, paired t-test). This advantage diminished at higher coherence levels (±30%, ±50%), where performance approached ceiling for both directions.

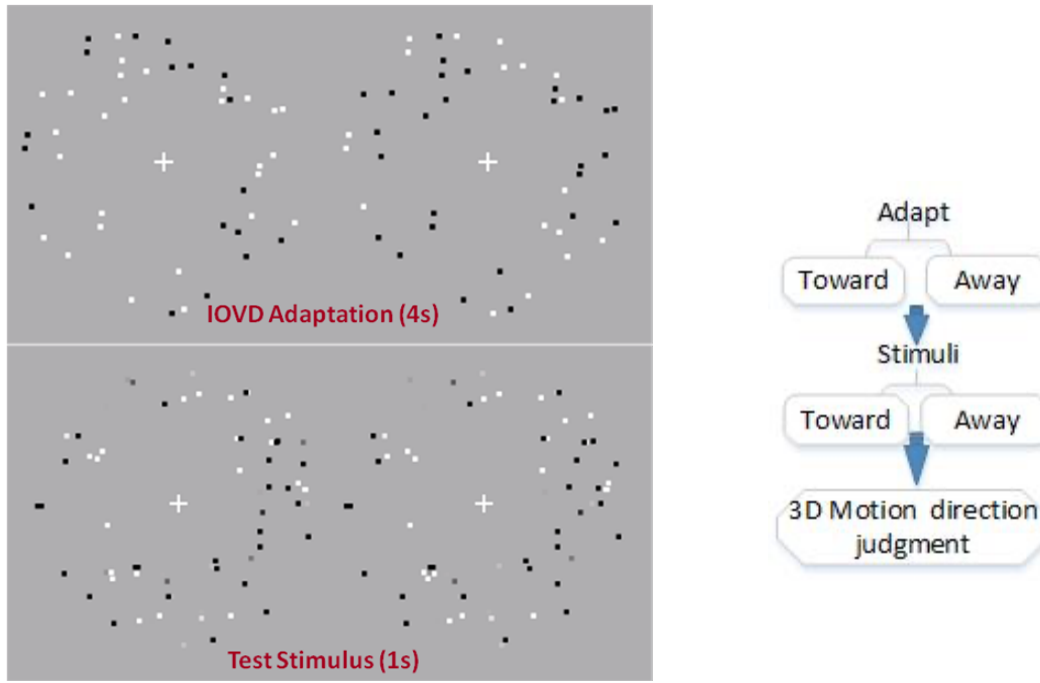

Figure 1: A) Example of IOVD 4 s adaptation and 1 s test stimulus during which observers are required to discriminate motion direction and answer. B) Adaptation session schedule for the MAE generation.

This baseline asymmetry in motion sensitivity is consistent with previous findings showing greater sensitivity for detecting objects moving toward versus away from the observer (Perrone, 1986; Shirai & Yamaguchi, 2004), and for discriminating crossed versus uncrossed disparities (Mustillo et al., 1988).

### 3.2.2 Asymmetry in Motion Aftereffect Magnitude

Figure 2 shows the psychometric functions for the unadapted condition and the two adaptation conditions. The fitted logistic parameters for each condition were as follows:

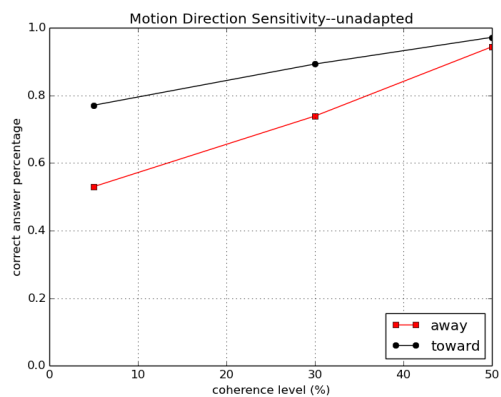

Figure 2. Motion direction sensitivity for the directions of motion-toward and motion-away from observers

The unadapted curve showed a slight negative bias (α = -0.072), reflecting the baseline directional sensitivity advantage for motion-toward noted above. Using this as the reference point, we found that adaptation to motion-toward shifted the psychometric function rightward by approximately 0.177 (from -0.072 to 0.105), indicating that approximately 17.7% coherent motion-toward was needed to null the motion-away aftereffect. Conversely, adaptation to motion-away shifted the function leftward by approximately 0.099 (from -0.072 to -0.171), indicating that approximately 9.9% coherent motion-away was needed to null the motion-toward aftereffect.

This asymmetry in MAE magnitude suggests that adaptation to motion-toward produces a stronger aftereffect than adaptation to motion-away, even after accounting for the baseline bias. It is worth noting, however, that while the difference between adaptation conditions was consistent across observers, the confidence intervals for the individual parameter estimates were relatively wide, indicating substantial variability.

### 3.3 Discussion

The results of Experiment 1 demonstrate two key findings. First, there is a baseline advantage in sensitivity for detecting motion toward versus away from the observer, consistent with previous literature on directional anisotropies in stereoscopic vision. Second, there is an asymmetry in the magnitude of motion aftereffects, with adaptation to motion-toward producing stronger aftereffects than adaptation to motion-away.

This asymmetry in MAE strength could reflect fundamental differences in the neural processing of approaching versus receding motion. The visual system may allocate more resources to processing motion-toward, potentially as an evolutionary adaptation for detecting approaching objects that might pose a threat or require rapid action. This enhanced processing could lead to stronger adaptation effects and consequently stronger aftereffects.

It is important to note that our analysis focused on shifts in the point of subjective equality (α parameter) rather than changes in the slope of the psychometric function (β parameter). While there were some differences in slope between conditions, these were relatively minor and did not show a consistent pattern across observers.

# 4. Experiment 2: Effect of Induced Eye Vergence on Motion Aftereffect

## 4.1 Methods

Experiment 2 investigated how induced vergence eye movements affect the magnitude and directional asymmetry of MAE in depth. The setup was identical to Experiment 1, with one critical difference: the fixation cross now oscillated in depth with a sinusoidal pattern, inducing changes in vergence eye movements.

The binocular disparity of the fixation cross varied according to the equation (2):

$$2\arctan\left(\frac{IOD * a * \sin(\omega t)}{0.7 * (0.7 + a * \sin(\omega t))}\right) \quad (2)$$

where «IOD» is the inter-ocular distance of the observer, $a$ – *amplitude of sine* and $\omega$ the pulse of the sine function which is $2\pi F$, where « F » is frequency of sinusoidal oscillations. Vergence demand ranges in virtual space between 1.14 (0.87 m) to 2.14 MA (0.47m) at 0.25 Hz. These parameters were adjusted from the previous studies in order to facilitate achievement of the visual task.

This oscillation induced both changes in absolute disparity (through vergence changes) and relative disparity between the fixation cross and the surrounding dots. Critically, the relative disparity changes could enhance or diminish the effective motion signal depending on the phase relationship between the cross motion and the dot motion.

Each observer completed 96 trials (16 trials × 6 coherence levels) in the unadapted session (approximately 5.6 minutes) and 160 trials (16 trials × 10 coherence levels) in each adaptation session (approximately 20 minutes each). The increased number of trials per condition compared to Experiment 1 was to account for the additional variability introduced by the moving fixation cross.

## 4.2 Results

### 4.2.1 Effect of Cross Motion on Baseline Perception

We first examined how the oscillating fixation cross affected baseline motion perception in the unadapted condition. Figure 3 compares the psychometric functions for three conditions: static fixation cross (from Experiment 1), oscillating cross with initial motion-toward, and oscillating cross with initial motion-away.

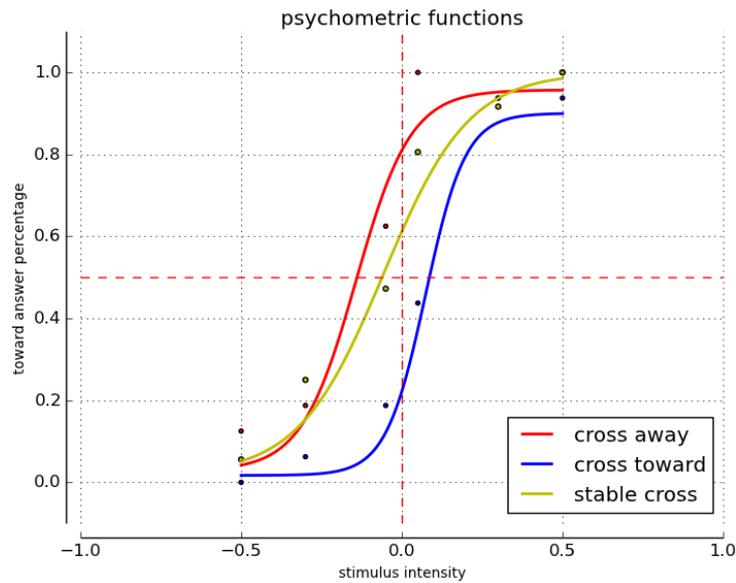

Figure 4. The comparison of the un-adapted tests. Red curve corresponds to the fitted logistic function in case of the cross motion away. The green line is the fitted logistic function for stable cross. The blue curve represents the fitted logistic function for cross moving toward.

The oscillation of the fixation cross systematically biased motion perception. When the cross moved toward the observer, dots appeared more likely to move away, shifting the psychometric function rightward. Conversely, when the cross moved away, dots appeared more likely to move toward, shifting the function leftward. These shifts can be attributed to changes in relative disparity between the cross and the surrounding dots, creating a contrast effect in perceived motion direction.

Interestingly, the magnitude of this effect was asymmetrical, with the toward-moving cross having a larger impact on motion perception than the away-moving cross. This asymmetry further supports our finding from Experiment 1 that the visual system is more sensitive to motion-toward than motion-away.

**4.2.2 Effect of Vergence on Motion Aftereffect Magnitude and Asymmetry**

Figure 4 shows the psychometric functions for the unadapted and adaptation conditions with the oscillating fixation cross. The fitted logistic parameters were:

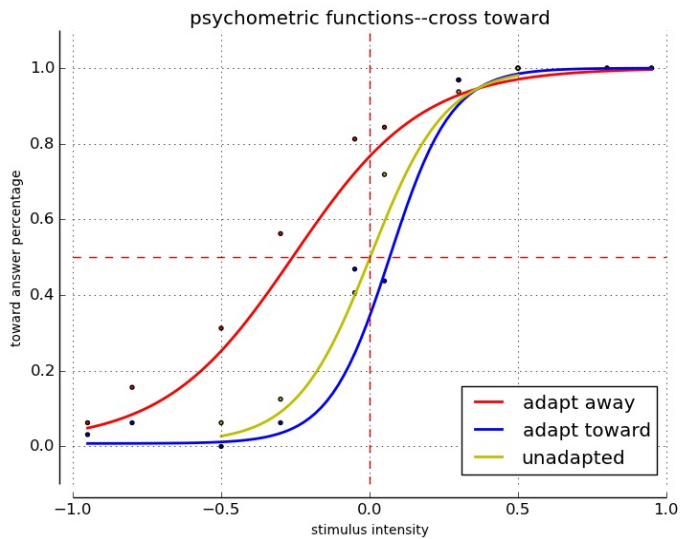

Figure 5. The motion-in-depth perceptions in case of implementing sinusoidal disparity on the fixation cross. Yellow curve is fitted in the case of un-adapted test(α= 0.03, CI95 = [-0.095, 0.112]; ß-1 = 0.126 (CI95 = [0.036, 0.204]), red curve is is fitted in the case adaptation motion-away (α= -0.225, CI95 = [-0.426, -0.093]; ß-1 = 0.217 (CI95 = [0.113, 0.318]), blue curve is is fitted in the case adaptation motion-toward (α= 0.069, CI95 = [-0.021, 0.169]; ß-1 = 0.103 (CI95 = [0.012, 0.159])

The most striking finding was the reversal of the directional asymmetry observed in Experiment 1. With the oscillating fixation cross, adaptation to motion-away produced a stronger MAE (leftward shift of approximately 0.255, from 0.03 to -0.225) than adaptation to motion-toward (rightward shift of approximately 0.039, from 0.03 to 0.069).

This dramatic change in the pattern of results suggests a complex interaction between adaptation direction and vergence state. To understand this interaction, we analyzed the relative disparity changes between the fixation cross and the RDS during the adaptation period (Figure 5).

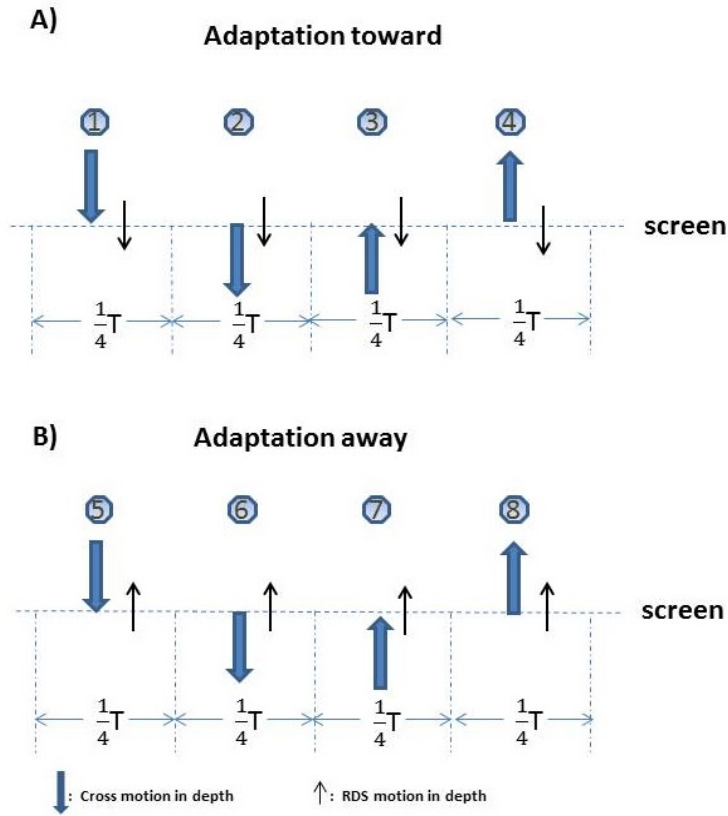

Figure 6. The relative disparity changment during 4s adaptation. (A) 4 relative positions in depth of the cross and RDS during 4s adaptation motion-toward (B) 4 relative positions in depth of the cross and RDS during 4s adaptation motion-away.

During the 4-second adaptation period, the relative disparity between the cross and the RDS varied systematically. The most significant MAE contribution likely occurred during periods when the cross and RDS moved in opposite directions, creating large relative disparity changes. In these situations, the vergence state (convergence vs. divergence) interacted with the adaptation direction to modulate the strength of the adaptation effect.

Specifically, during adaptation to motion-away, the strongest adaptation occurred during periods of eye convergence (when the cross moved toward). Conversely, during adaptation to motion-toward, the strongest adaptation occurred during periods of eye divergence (when the cross moved away). Our results suggest that adaptation during convergence produced stronger MAEs than adaptation during divergence, leading to the reversal of the directional asymmetry observed in Experiment 1.

### 4.3 Discussion

Experiment 2 demonstrated that induced vergence eye movements substantially affect both baseline motion perception and the magnitude and directional asymmetry of MAE in depth. The oscillating fixation cross introduced systematic biases in motion perception through changes in relative disparity between the cross and the surrounding dots.

More importantly, the induced vergence reversed the directional asymmetry in MAE magnitude observed in Experiment 1. With a static fixation (Experiment 1), adaptation to motion-toward produced stronger MAEs than adaptation to motion-away. With an oscillating fixation (Experiment 2), adaptation to motion-away produced stronger MAEs than adaptation to motion-toward.

This reversal suggests that the state of vergence during adaptation is a critical factor in determining the strength of the resulting MAE. Specifically, adaptation during convergence appears to produce stronger MAEs than adaptation during divergence, regardless of the direction of the adapting motion. This finding highlights the importance of extra-retinal signals in motion adaptation and challenges purely retinal accounts of MAE in depth.

It is worth noting that observers reported greater difficulty in Experiment 2 compared to Experiment 1, likely due to the additional cognitive and perceptual demands of tracking the oscillating fixation cross while judging motion direction. This increased difficulty might have contributed to the variability in our results, as reflected in the wider confidence intervals for the fitted parameters.

## 5. General Discussion

### 5.1 Directional Asymmetry in Motion-in-Depth Perception

Our findings provide evidence for a directional asymmetry in both the perception of and adaptation to motion in depth. In Experiment 1, we observed a baseline advantage for detecting motion toward versus away from the observer, consistent with previous research showing enhanced sensitivity for crossed versus uncrossed disparities (Mustillo et al., 1988) and for expanding versus contracting optic flow (Shirai & Yamaguchi, 2004).

This directional asymmetry may reflect an evolutionary adaptation of the visual system to prioritize approaching objects, which often require rapid responses to avoid collision or to prepare for interaction. Neural evidence supports this interpretation, with studies showing specialized populations of neurons in visual cortex that respond preferentially to motion toward the observer (Perrone, 1986; Wang & Yao, 2011).

The asymmetry extended to motion adaptation, with stronger MAEs following adaptation to motion-toward than motion-away in Experiment 1. This asymmetry could indicate that the neural populations processing motion-toward are more numerous or more plastic than those processing motion-away, leading to stronger adaptation effects.

## 5.2 Role of Eye Vergence in Motion-in-Depth Perception

Experiment 2 revealed a complex interaction between vergence eye movements and motion adaptation. The oscillating fixation cross not only introduced systematic biases in baseline motion perception through relative disparity changes but also reversed the directional asymmetry in MAE magnitude.

This finding underscores the importance of extra-retinal signals in motion-in-depth perception, consistent with previous research showing that vergence eye movements modulate perceived motion speed (Nefs & Harris, 2007) and contribute to accurate depth estimation (Welchman et al., 2009). Our results extend this work by demonstrating that vergence state also affects the strength of motion adaptation.

The reversal of the directional asymmetry suggests that the combination of retinal motion signals and vergence state creates a complex interaction that modulates the effectiveness of motion adaptation. Specifically, adaptation during convergence (when eyes are rotating inward to focus on a closer object) appears to produce stronger adaptation effects than adaptation during divergence (when eyes are rotating outward to focus on a more distant object).

This pattern might reflect the ecological statistics of natural viewing, where objects moving toward often require convergence to maintain fixation, creating a stronger association between convergence and approaching motion in the visual system. Alternatively, it could reflect fundamental differences in the neural mechanisms underlying convergence versus divergence movements, with convergence potentially engaging more neural resources or different neural populations than divergence.

## 5.3 Implications for Models of Motion-in-Depth Processing

Our findings have several implications for models of motion-in-depth processing. First, they suggest that models should incorporate directional asymmetries, with enhanced processing for motion-toward compared to motion-away. Second, they highlight the need to consider the influence of vergence eye movements and the interaction between retinal and extra-retinal signals in motion perception.

Current models often focus exclusively on the retinal cues to motion in depth (CD and IOVD mechanisms) without considering how eye movements modulate the processing of these cues. Our results suggest that a complete model must incorporate both retinal and extra-retinal signals, as well as their interactions.

Furthermore, the asymmetrical effects we observed suggest that separate neural populations might be responsible for processing motion toward versus away from the observer, rather than a single population encoding motion in depth bidirectionally. This perspective aligns with recent neuroimaging studies showing distinct patterns of activation for approaching versus receding motion (Likova & Tyler, 2007).

## 5.4 Limitations and Future Directions

Several limitations of the current study should be acknowledged. First, our sample size was relatively small (three observers), potentially limiting the generalizability of our findings. However, the consistent pattern of results across observers and the use of multiple measurements per condition strengthen the reliability of our conclusions.

Second, the oscillating fixation cross in Experiment 2 introduced multiple factors simultaneously: changes in absolute disparity, changes in relative disparity, and induced vergence eye movements. This complexity makes it challenging to isolate the specific contribution of each factor to the observed effects. Future studies could address this limitation by systematically manipulating each factor independently.

For example, a future experiment could compare a condition with actual vergence eye movements (oscillating fixation cross) to a condition with simulated oscillation where the entire stimulus volume oscillates while vergence remains fixed. This comparison would help distinguish the effects of vergence eye movements from those of changing disparities.

Additionally, eye-tracking measurements could provide direct evidence of vergence movements and their relationship to the observed perceptual effects. Such measurements would allow for more precise correlations between vergence state and adaptation strength.

Finally, neuroimaging studies could help identify the neural populations involved in processing motion toward versus away from the observer and how these populations are modulated by vergence eye movements. Such studies would provide valuable insights into the neural basis of the behavioral effects observed in the current study.

## 6. Conclusion

This study investigated two key aspects of motion aftereffect in depth: directional asymmetry and the influence of vergence eye movements. We found evidence for a directional asymmetry in both motion perception and adaptation, with enhanced sensitivity for motion toward versus away from the observer and stronger MAEs following adaptation to motion-toward than motion-away when fixation was static.

More importantly, we demonstrated that induced vergence eye movements substantially affect motion adaptation, reversing the directional asymmetry observed with static fixation. This finding highlights the complex interaction between retinal motion signals and extraretinal vergence signals in the perception of motion through depth.

Together, these results enhance our understanding of the mechanisms underlying motion-in-depth perception and challenge purely retinal accounts of motion adaptation. They suggest that a complete model of motion-in-depth processing must incorporate both directional asymmetries and the modulatory influence of eye movements.